\documentstyle[twocolumn,prl,aps]{revtex}
\tighten

\begin{document}

\twocolumn[\hsize\textwidth\columnwidth\hsize\csname
           @twocolumnfalse\endcsname

\bibliographystyle{unsrt}
\draft

\title{
%
%
        {\small{\rm{{\hspace{10 cm}}UPR-0546-T/Rev}}}\\
        {\mbox{ }}\\
%
%
Non-Extreme and Ultra-Extreme Domain Walls and Their Global Space-Times}

%
%
\author{Mirjam Cveti{\v{c}}\cite{mirjam}
\and
{\mbox{ }}Stephen Griffies\cite{steve}
\and
{\mbox{ }}and
{\mbox{ }}Harald H. Soleng\cite{hjem}   }

\address{Department of Physics, University of Pennsylvania,
209 South 33rd Street, Philadelphia, Pennsylvania 19104-6396,  U.~S.~A.}


\maketitle

\begin{abstract}
Non-extreme walls (bubbles with two insides)
and ultra-extreme walls (bubbles of
false vacuum decay)
are discussed.
Their respective
energy densities are
higher and lower than that of
the corresponding
extreme  (supersymmetric), planar domain
wall. These singularity free space-times
 exhibit non-trivial causal
structure analogous to certain
non-extreme black holes.
We focus on  anti-de~Sitter--Minkowski
walls and comment on
Minkowski--Minkowski walls with trivial extreme limit,
as well as walls adjacent to de~Sitter space-times with
no extreme limit.
\end{abstract}

\pacs{PACS numbers: 04.20.-q, 04.65.+e, 11.30.Pb}

\vskip2pc]

Domain walls can form as topological defects
in the early Universe in theories with isolated
minima of the matter potential\cite{Review} or
as boundaries of true vacuum bubbles nucleating
in a false vacuum
\cite{Coleman}.
The induced space-times of domain walls
provide a fertile ground to study
globally non-trivial space-times
without singularities.  Here we discuss
certain domain walls which are natural generalizations
of the planar, extreme
domain walls
with energy density
$\sigma_{\text{ext}}
\equiv
\sigma_{\text{susy}}$,
that separate isolated
supersymmetric vacua
\cite{CQR,CGRI,CG,CGII,CDGS,Gibb}.
The non-extreme wall ($\sigma_{\text{non}}>\sigma_{\text{ext}}$)
corresponds to a bubble with  two insides;
i.e., each side of the wall is
inside a bubble.
The ultra-extreme wall ($\sigma_{\text{ultra}}
<\sigma_{\text{ext}}$) corresponds
to the false vacuum decay
\cite{Coleman,CD}
tunneling bubble
\cite{BKT}.
We explain the relation between the
non- and ultra-extreme domain wall bubbles \cite{BKT} and the
supersymmetric extreme walls
\cite{CQR,CGRI,CG,CGII,CDGS,Gibb}.
The interesting global structure
of the solutions with analogies to certain black holes
is pointed out.

We choose to describe
the gravitational field in the rest frame of the wall,
i.e.\ we use comoving coordinates of observers
sitting on the wall. Hence, the
wall is placed at a fixed
$z$-coordinate, and the metric is
static in the $(t,z)$-directions transverse to the wall.
The metric is assumed to be homogeneous
and isotropic in the $(\varrho,\phi)$ surfaces parallel to the wall
\cite{FOOTV}.
Since the extrinsic  curvature is independent of
the wall's proper time,
one can show \cite{CGSII}
that the metric is
\begin{equation}
ds^2=A(z)\left( dt^2 - dz^2 - \beta^{-2} \; \cosh^2\!\beta t
\; d\Omega^2_{2}\right)\; ,
\label{metne}
\end{equation}
with $A(z)>0$
and $d\Omega^2_{2}\equiv [1-(\beta \varrho)^2]^{-1}d(\beta \varrho)^2
+ (\beta \varrho)^2d\phi^2$.
In the extreme limit, $\beta \rightarrow 0$, the $(\varrho,\phi)$ surface
becomes a plane with $\varrho$ and $\phi$ planar polar coordinates.
When
$\beta\neq 0$, the $(\varrho,\phi)$ hyperspace is the surface of a
three-dimensional
sphere, that is, its topology is ${\bf S}^2$ \cite{Gibb}.
In this case the coordinate
$\varrho = \beta^{-1} \sin\theta$ is compact.
The scalar curvature of the
spatial ${\bf S}^{2}$ is $2\beta^2 A(z)^{-1}( \cosh\beta t)^{-2}$.
The constant $z$ section with $\beta\neq 0$
is (2+1)-dimensional de~Sitter space-time (dS$_{3}$),
which has the topology ${\bf R}(\text{time})\times
{\bf S}^2(\text{space})$~\cite{desitterinside}.  dS$_{3}$
is completely covered by the coordinates $(t,\theta,\phi)$.
Indeed, geodesic completeness for this (2+1)-dimensional
space-time
requires the use of the compact spatial section
in the directions parallel to the wall~\cite{HE}.
The novel issues of geodesic completeness in
the (1+1)-dimensional
space-time transverse to the wall, the $(t,z)$-directions, will
be addressed in this Letter.

The extreme walls ($\beta=0$) induce a static, conformally flat
space-time \cite{CGRI,CG,LINET} classified in Ref.\ \cite{CG}:
two types of AdS$_4$--AdS$_4$ walls
(Type II and Type III) and an
AdS$_4$--M$_4$ wall (Type I) where AdS$_{4}$ and M$_{4}$ denote
the type
of asymptotic geometry away from the wall, i.e.\
anti-de~Sitter
and Minkowski space-time, respectively.
Here, we focus on
space-times which
are
asymptotically M$_{4}$ in the direction transverse to
the wall (Type I and its generalizations), and
comment on
the other possibilities at the end.

The extreme Type I wall energy density,  $\sigma_{\text{ext}}$,
and the conformal factor, $A(z)$, are
\begin{equation}
\left.
\begin{array}{ccl}
\sigma_{\text{ext}}&=&2\kappa^{-1}\alpha  \\
A(z)&=&\left\{ \begin{array}{ll}
  ( \alpha z-1)^{-2}\;\;
&
z <  - w\\
1 &
z >  + w\; ,
              \end{array}
      \right.
\end{array}
\right.
\label{solI}
\end{equation}
where, without loss of generality, the wall
is centered
at $z=0$.
$2w> 0$ is the width of the wall and $\kappa\equiv 8\pi G$.
The M$_{4}$  side is chosen
to be at $z > w$,
and the cosmological constant
$\Lambda\equiv -3\alpha^2$
with $\alpha\geq 0$ on the $z<w$ side.
The
{\em horo-spherical\/}
coordinates used on the AdS$_{4}$ side are
discussed in\cite{CG,Gibb,Griffies}. In the supersymmetric model
the fields are governed
by coupled first order rather than second order differential equations,
thus  allowing  for a straightforward solution of the field equations
for any thickness of the wall \cite{CGRI,CG}.
The coordinates of the metric (\ref{metne})
are not geodesically complete in the $(t,z)$ directions.
Geodesic extensions have been provided
with emphasis on the Type I
walls in Ref.\ \cite{CDGS} and
Type II walls in Ref.\ \cite{Gibb}.

We now discuss walls with $\beta > 0$.
Primarily we describe
infinitely thin walls, $w=0$,  and thus employ Israel's
formalism of singular hypersurfaces \cite{ISR}.
Nevertheless, we motivate our analysis from
an underlying scalar field theory and  mention generic
thick wall results where appropriate.
Israel's  matching conditions at the wall are
$\kappa S^{\mu}_{\;\;\nu} \equiv-[K^{\mu}_{\;\;\nu}]^{-}
+\delta^{\mu}_{\;\;\nu}[K]^{-}=
2
(\delta^{\mu}_{\;\;\nu}+n^{\mu}n_{\nu})
[K^{t}_{\;\;t}]^{-}
$.
Here  $K^{i}_{\;j} \equiv - n^{i}_{\; ;\, j}$
is the wall's extrinsic curvature,
and $n^{\nu}=\pm A^{-1/2}\delta^{\nu}_{\;\;z}$
is the space-like {\em unit\/} normal orthogonal to the wall's
four velocity $u^{\nu} = A^{-1/2}\delta^{\nu}_{\;\;t}$.
$[K]^{-} \equiv  K_{(z \rightarrow  0^+)} -
K_{(z \rightarrow 0^-)} $ signifies  the
discontinuity of the extrinsic curvature
at the wall, and the Lanczos surface energy-momentum tensor
$\kappa S^{i}_{\;j} = \sigma \delta^{i}_{\;j}$ is
of the domain wall form \cite{VILENKIN}.
The  sign ambiguity of $n^{\mu}$ is
resolved by demanding a positive energy density for the wall
and by using the underlying scalar field theory to identify the
wall with a kink-like source.
Then, Einstein's field equations and Israel's matching method
yield two kinds of solutions with energy density
and conformal factor
\begin{equation}
\left.
\begin{array}{ccl}
\sigma^{\text{non}}_{\text{ultra}}
& =& 2 \kappa^{-1}
[ (\alpha^{2} + \beta^{2} )^{1/2} \pm \beta ]\\
A(z) &= & \left\{ \begin{array}{ll}
\beta^2\alpha^{-2}[\sinh(\beta z -\beta z')]^{-2}\;\;
 &
z < 0 \\
e^{\mp 2\beta z}
 &
z > 0\; ,
               \end{array}
               \right.
\end{array}
\right.
\label{solII}
\end{equation}
where  $e^{2\beta z'} \equiv [\alpha^2 + 2\beta^{2} +  2 \beta
 (\beta^{2} + \alpha^2 )^{1/2}]/\alpha^2 \equiv \eta\ge 1$
is  determined by $A(0) \equiv 1$.

The upper sign solution  of Eq.\ (\ref{solII})
represents a {\em non-extreme wall.\/}
In the non-extreme wall region
the
potential barrier associated with the scalar field
is larger
than in the corresponding extreme domain wall \cite{FOOTVIS},
which implies
that $\sigma_{\text{non}} > \sigma_{\text{ext}}$,
and that $A(z)$
falls off on the $M_4$ side.
Within $N=1$ supergravity theory,
it can be shown that such a wall can be realized
as a wall interpolating between a
supersymmetric  M$_{4}$ vacuum and an AdS$_{4}$
vacuum with
supersymmetry spontaneously broken.

At  $t=0$  the bubble has a radius $\beta^{-1}$ which then increases as
as $\cosh\beta t$. Additionally, since the
radius of the bubble $\beta^{-1} A(z)^{1/2} \cosh\beta t$ decreases
as we move spatially away from the bubble
in both $z$ directions,  observers on {\em both\/} sides are
{\em inside\/} the bubble.

The non-extreme walls exhibit cosmological horizons
on both the AdS$_4$ and M$_4$ sides. Namely, a particle
with energy per unit mass $\epsilon \ge 1$,
freely falling at constant
$\theta$ and $\phi$ in the $z \rightarrow \mp \infty$-direction,
has a finite proper time
$ \tau =
\alpha^{-1}(\arcsin\{ [ 1 + (\epsilon \alpha  / \beta)^{2} ]^{-1/2}
(\eta + 1)/ (\eta - 1) \}
- \arcsin [ 1 + (\epsilon \alpha  /
\beta)^{2} ]^{-1/2})$
and $\tau = \beta^{-1} [ \epsilon - (\epsilon^{2} - 1)^{1/2}]$, respectively.
As $\beta \rightarrow  0$,
the
cosmological horizon
on the AdS$_{4}$ side becomes a Cauchy horizon (as in the extreme wall
space-time) with
$\tau = \alpha^{-1} \arcsin(1/ \epsilon) $, while the
M$_{4}$ side becomes geodesically complete \cite{CDGS}.

To investigate geodesically complete space-times for the non-extreme
walls, we transform the metric (\ref{solII}) to the inertial spherical
M$_{4}$ and {\em Einstein cylinder}
AdS$_{4}$ coordinates on the respective sides.
Introducing the radial Rindler coordinates
$\underline{t} = \beta^{-1} e^{- \beta z} \sinh{\beta t}$
and
$\underline{r} = \beta^{-1} e^{- \beta z} \cosh{\beta t}$
brings the line element on the M$_{4}$ side to the spherically
symmetric form
$ds^{2} = d\underline{t}^{2} - d\underline{r}^{2} - \underline{r}^{2}
d\Omega_{2}^{2}$.
The $(\underline{t}, \underline{r}, \theta, \phi)$ coordinates
define an intertial frame in which the bubble at $z=0^{+}$
lives on the hyperbolic trajectory
$\underline{r}^{2} - \underline{t}^{2} =
-\tan(u'/2) \tan(v'/2) = \beta^{-2}$ with constant
acceleration $\beta$; i.e.\ a Rindler trajectory \cite{BD}. Here
$u',v' = 2\tan^{-1}[\beta(\underline{t} \mp \underline{r})]$
are the usual compact null coordinates.
On the AdS$_{4}$ side, we map to the spherically
symmetric {\em Einstein cylinder\/} coordinates \cite{HE}.
This transformation is done in three steps:
($i$) $\ln \Xi = \beta(z-z')$.
($ii$) Radial Rindler transformation:
$T = \Xi \sinh\beta t$ and $R = \Xi \cosh\beta t$.
($iii$) Compact time-like and radial coordinates:
$T \pm R = \tan[(t_{c} \pm \psi)/2] $.
The line element on the AdS$_{4}$ side ($z<0$) becomes
$ds^{2} = (\alpha \cos \psi)^{-2}(dt_{c}^{2} - d\psi^{2} - \sin^{2}\psi
d\Omega_{2}^{2})$,
where $-\pi \le t_{c} \pm \psi  \le \pi$ and $0 \le \psi \le \pi/2$.
The center of symmetry is at $\psi = R = 0$.
The bubble at $z=0^{-}$ again lives on a hyperbolic trajectory
$R^{2} - T^{2} =
-\tan[(t_{c} - \psi) / 2] \tan[(t_{c} + \psi) / 2]
= \eta^{-1}$.

The $(t,z)$-chart is an  interpolating map which covers
the space-time on both sides of the non-extreme wall region.
To complete the space-time,
we extend onto pure M$_{4}$ and  AdS$_{4}$
on the respective sides, as shown in
Fig.\ \ref{fig1}.
On the AdS$_4$ side, one may consider a symmetric, periodic
extension yielding a
lattice structure of walls.  The
Penrose diagram for this extension
bears remarkable similarities to the one of
a non-extreme
($ m^{2} G   > e^{2}$)
Reissner-Nordstr{\"o}m (RN)
black hole; however, {\em without\/}
singularities. The endpoints of the wall trajectories
are on the affine boundary of
AdS$_{4}$\cite{CDGS,Gibb,AIS} and M$_{4}$
and thus are not probed.  The AdS$_4$ ({\bf A})  and
M$_4$ ({\bf M}) diagrams are
linked to each other at the wall regions.

The analogy between the space-time of the walls and that of
black holes goes further.
On the
AdS$_{4}$ side
of the non-extreme walls, the metric in the $(t,z)$-directions is
identical to that of the $(t,r)$-directions near the event horizon of
non-extreme black holes.
For the RN system in its $(t,r)$-section,
we have the line element
$ds^{2} = a(\rho) dt^{2} - a(\rho)^{-1} d\rho^{2}$, where
$\rho\equiv r -  r_{+} \rightarrow  0^{+}$,
$a(\rho) = \rho (\rho + \Delta r) / r_{+}^{2}$,
$r_{\pm} = G [  m \pm (m^{2}- e^{2} G^{-1})^{1/2} ]$
and $\Delta r\equiv
r_{+} - r_{-}$. Defining $e^{ 2\beta z}
\equiv \eta \rho/(\rho +\Delta r)
$ along with $2\beta=\Delta r/r_+
^{2}$ and $ \alpha = 1 / r_{+}$,
brings  the above metric to the form
$ds^{2} = A(z)(dt^{2} - dz^{2})$
where $A(z)$ is the conformal factor on
the AdS$_4$ side of the wall
(Eq.\ (\ref{solII}) for
$z < 0$).
As one approaches the extreme RN limit
($m^{2} G \rightarrow  e^{2})$,
$\Delta r \rightarrow  0$, and $A(z)$ reduces to
that of the extreme domain
wall (Eq.\ (\ref{solI})  for $z < 0$).
Furthermore, on the M$_{4}$ side of the wall,
the metric in the $(t,z)$-directions (Eq.\ (\ref{solII}) for
$z > 0$) corresponds to the $(t,r)$-directions of the
Schwarzschild horizon \cite{BD}: $ds^2=a(\rho)dt^2-
a(\rho)^{-1} d\rho^{2} $  with $a(\rho)=\rho(2 m G)^{-1}$
and
$\rho\equiv r - 2 m G
\rightarrow 0$.  Here
we set $\rho\equiv (2mG) e^{-2 \beta  z}$
and $\beta^{-1}  \equiv 4 m G$.

The lower sign solution of  Eq.\ (\ref{solII})
describes an {\em ultra-extreme wall.\/}
For these walls
the potential barrier associated
with the scalar field is smaller than that
of the extreme walls
\cite{FOOTVIS},
which means $\sigma_{\text{ultra}} < \sigma_{\text{ext}}$ and the metric
blows up on
the M$_4$ side.
Ultra-extreme  walls exhibit the same causal structure
on the AdS$_{4}$ side as the non-extreme wall.
However, the M$_{4}$ side is geodesically complete
in the $(t,z)$-coordinates.
The M$_{4}$ side is the complement of the M$_{4}$ side of the
non-extreme wall (see Fig.\ \ref{fig2}).  The two diagrams are linked
at the  wall region.
The regions bounded by the curved trajectory
of the wall and the
null infinities are covered by the
$(t,z)$-coordinates.

The Minkowski side is on the {\em outside\/} of the
ultra-extreme bubble because the
radius $\beta^{-1} A(z)^{1/2} \cosh\beta t$
increases with $z$
on the $z>0$
side. On the AdS$_{4}$ side; however,  the radius
decreases away from the wall, and thus
AdS$_{4}$ is on the
{\em inside\/} just as for the non-extreme solution.
Since $\sigma_{\text{ultra}} < 2\kappa^{-1}\alpha$,
i.e.\ below the Coleman-De~Luccia bound \cite{CD,CGR},
the ultra-extreme
solution for $t\geq 0$ describes the classical evolution of
a bubble \cite{BKT} of true vacuum
created by the quantum tunneling process of false vacuum decay
\cite{Coleman,CD}.
At $t=0$ the bubble is formed with
radius $\beta^{-1}$, expands as
$\cosh\beta t$, and
inevitably hits
all time-like observers on the M$_{4}$ side.
If there were no Cauchy
horizons, the
AdS$_{4}$ side would collapse to a singularity \cite{AbbCol}.
However, as shown in Fig.\ \ref{fig1}
there {\em are\/} Cauchy horizons on the AdS$_{4}$ side.
Thus,
the conclusion of Ref.\
\cite{AbbCol} that the AdS$_{4}$ space collapses
does not apply.

For completeness, we also give results
for the Israel matching of thin AdS$_{4}$--AdS$_{4}$ walls.
Type II walls \cite{CG,LINET},
with $\sigma_{\text{ext}} = 2\kappa^{-1} (\alpha_1+\alpha_2)$,
have a unique non-extreme
counterpart, with $\sigma_{\text{non}}=
 2 \kappa^{-1}[ ( \alpha_{1}^{2} + \beta^{2} )^{1/2} +
( \alpha_{2}^{2} + \beta^{2} )^{1/2}  ]$.
Type III walls \cite{CG}, with $\sigma_{\text{ext}}= 2 \kappa^{-1}
(\alpha_{1} - \alpha_{2})$,
have a unique ultra-extreme
counterpart  with $\sigma_{\text{ultra}}=
2 \kappa^{-1} [ (\alpha_{1}^{2} + \beta^{2} )^{1/2} -
( \alpha_{2}^{2} + \beta^{2} )^{1/2} ]$.
Both walls have analogous solutions
for the metric coefficient $A(z)$ and the
geodesic extensions~\cite{CGSII}.

The M$_{4}$--M$_{4}$
walls~\cite{VILENKIN}
and their geodesic extensions
correspond to the $\alpha \rightarrow  0$ limit
of the non-extreme AdS$_{4}$--M$_{4}$
walls with the metric (\ref{metne}).
In  this limit,  Eq.\ (\ref{solII})
reduces to $\sigma_{\text{non}} = 4 \kappa^{-1}\beta$  and
$A(z)=e^{-2 \beta |z|}$ for $|z|>0$.
Such walls have a trivial extreme limit  $\beta \rightarrow  0$
with  $\sigma_{\text{non}}= 4\kappa^{-1}\beta \rightarrow  0$.
This is analogous  to the case of the
Schwarzschild space-time, which admits supersymmetry
only in the trivial case of vanishing mass.

There are also walls separating de~Sitter space (dS$_{4}$) from other vacua
\cite{Sato,Blau,BKT2}.
An AdS$_{4}$--dS$_{4}$ wall ($\Lambda_{2}\equiv +3\alpha_{2}^{2}$)
has two wall solutions  with
$\sigma_{\pm} = 2 \kappa^{-1}
[ (\alpha_{1}^{2} + \beta^{2} )^{1/2} \pm
(-\alpha_{2}^{2} + \beta^{2} )^{1/2} ]$,
whereas a dS$_{4}$--dS$_{4}$ wall has the two solutions
$\sigma_{\pm} = 2 \kappa^{-1}
[(- \alpha_{1}^{2} + \beta^{2} )^{1/2} \pm
(-\alpha_{2}^{2} + \beta^{2} )^{1/2} ]$. M$_{4}$--dS$_{4}$ walls
correspond to the special case $\alpha_{1}= 0$.
None of these walls have an extreme limit ($\beta \rightarrow 0$)  since
$\beta$ cannot
be smaller than $\alpha$ of the dS$_{4}$ space(s) \cite{Sato}.
Additionally, they separate vacua which are unstable
to quantum tunneling \cite{CD} except
in the fine-tuned
dS$_{4}$--dS$_{4}$
case
$\alpha_{1} = \alpha_{2}=\alpha$ with $\sigma =
4 \kappa^{-1} (-\alpha^{2} + \beta^{2})^{1/2}$.

Local and global properties of exact domain wall
solutions have been analyzed.
The non-extreme wall corresponds to a bubble with two insides.
Its
energy density is bounded from below by the one of the extreme wall, which
is a planar supersymmetric configuration.
Since the energy density of the extreme domain wall
is equal to the
Coleman-De~Luccia bound~\cite{CD}, supersymmetry provides
a lower bound
\cite{KalGib} for a
non-extreme domain wall.
On the other hand, the ultra-extreme wall,
which has energy density lower than
the one of the extreme wall, corresponds
to the classical evolution of a bubble \cite{CD,BKT,BKT2} of true
AdS$_{4}$ created by
the decay of the false M$_{4}$ vacuum.

We would like to thank  R. Davis, G. Gibbons, G. Horowitz, G. Segr\` e,
and P. Steinhardt for useful discussions.
This work was supported in part by
U. S. DOE Grant No.\ DOE-EY-76-C-02-3071, SSC Junior Faculty
Fellowship (M. C.), NATO Research Grant No. 900-700 (M. C.),
Fridtjof Nansen Foundation Grant No.\
152/92 (H. H. S.),
Lise and Arnfinn Heje's Foundation
Ref.\ No.\ 0F0377/1992 (H. H. S.), and
by the Norwegian Research Council for Science and the
Humanities (NAVF), Grant No.\ 420.92/022 (H. H. S.).





\begin{figure}
\caption{Conformal diagram for the
non-extreme domain wall system.
The regions between the curved
lines, representing the wall, and the
dotted lines, representing the cosmological horizons,
are covered by the  $(t,z)$-chart.  We extend
across the cosmological horizons onto pure
AdS$_{4}$ ({\bf A}) and M$_{4}$ ({\bf M}).
The time-direction is vertical, and inside each bubble the
axis of symmetry represents the world line of the
bubble's center. Opposite points on the right and left side of this
axis represent antipodal points
($\theta \rightarrow \pi - \theta$ and $\phi \rightarrow \phi + \pi$).
The {\bf M} side is a cross-section of the hyperboloid of dS$_{3}$ as
embedded in M$_{4}$ (see Ref.\ [14] for the analogous case of dS$_{4}$).
The two diagrams are identified across the
wall region by revolving and rotating
the de~Sitter hyperboloid of the wall space-time
as embedded in Minkowski space, {\bf M},
around the AdS-cylinder, {\bf A},
and identifying adjacent points of the two wall
regions.
The dashed diagonals in the {\bf A} diagram correspond to the Cauchy
horizons of the AdS$_4$ space-time.
The wall on the AdS$_{4}$ side sweeps out a hyperbolic trajectory
over half the fundamental domain of pure AdS$_{4}$.
By introducing new walls above the original walls, instead of the spatial
infinities,
one can form an indefinite lattice structure.   Notice that the conformal
diagram of the lattice extension bears remarkable similarities to the one of
a non-extreme Reissner-Nordstr{\"o}m
black hole; however, {\em without\/}
singularities.}
\label{fig1}
\end{figure}


\begin{figure}
\caption{The M$_{4}$ side of
the ultra-extreme
domain wall.
The two sides represent anti-podal pieces of the spherically
symmetric space-time.
In this case the Minkowski side
corresponds to the outside of the de~Sitter hyperboloid
of Fig.\ 1.
The {\bf M} region (the M$_4$ side) is covered by $(t,z)$.
The AdS$_{4}$ side of the wall is
discussed in Fig.\ 1. The two diagrams are
glued to the AdS region of Fig.\ 1
by putting the AdS-cylinder in the hole of the
{\bf M} region and identifying across the wall regions.
Angular coordinates are suppressed as in Fig.\ 1.
}

\label{fig2}
\end{figure}


\begin{thebibliography}{100}
\bibitem{FOOTV} This is consistent with a scalar field source
$\Phi=\Phi(z)$, for which
$T^{\mu}_{\;\;\nu}=
 v\delta^{\mu}_{\;\;\nu}+  k (\delta^{\mu}_{\;\;\nu}+2n^{\mu}n_{\nu})$,
where $n^{\mu}\equiv\pm A^{-1/2}\delta^{\mu}_{\;\;z}$, and
$v$ and $k$ are the potential and kinetic energies, respectively.
\bibitem{FOOTVIS} Einstein's equations,
with a $z$ dependent scalar field,  are
$3H^{2} = 4 [3\beta^2+(k-v)A(z)]$,
$3\partial_{z} H= -2(2k+v)A(z)$, where $H\equiv \partial_z \ln A(z)$,
and $v$ and $k$ are given in \cite{FOOTV}.
For  $z < - w $, $k = 0$ and $v = -3 \kappa\alpha^2$
and for $z > w $, $k = 0$ and $v = 0$. However, inside
the wall ($|z|< w $),  $k >  0$.
\end{thebibliography}
\end{document}